# The next dimension: Digital holography for 3D interferometric scattering

Jaime Ortega Arroyo[1,*], Matz Liebel[2,*]

[1] Nanophotonic Systems Laboratory, Department of Mechanical and Process Engineering, ETH-Zürich, 8092 Zürich, Switzerland.

[2] Department of Physics and Astronomy, Vrije Universiteit Amsterdam, De Boelelaan 1100, Amsterdam, 1081 HZ, The Netherlands

* email: jarroyo@ethz.ch, m.liebel@vu.nl

**Abstract**

We provide detailed experimental guidelines for implementing digital holography in the context of high-sensitivity interferometric scattering (iSCAT) based nanosizing applications. Our approach relies on interferometry via the highly versatile off-axis implementation of digital holography, which offers key advantages over more traditional strategies. After a brief theoretical discussion of off-axis holography and its differences and similarities with iSCAT, typical experimental implementations and digital data-processing steps are presented. Key experimental parameters and strategies to achieve optimal performance are also highlighted. Following these experimental aspects, we focus on digital post-processing routines that enable digital refocussing and 3D particle tracking as well as pupil function aberration correction. We then conclude with a few examples highlighting the broad applicability of digital holography for nanosizing and particle characterisation applications as well as an outlook for future applications.

**Introduction,**

All-optical label free sizing and sensing approaches are highly relevant for addressing both fundamental as well as applied challenges. Applied, technologies such as nanoparticle tracking analysis or mass photometry are widely used in analytical labs for routine nano-characterisation[1,2]. Fundamentally, single-particle methods are prime candidates for answering biophysically relevant questions, especially when ensemble averaging masks the underlying dynamics[3–6]. Hisotrically, observations were often based on so-called darkfield observations where only light scattered by nano-objects of interest is detected[7–11]. However, it was soon realised that interferometric approaches offer key-advantages as they boost small scattering signals and exhibit favourable particle size-dependent signal scaling[12–15].

A very successful implementation, especially in the biophysics community, is interferometric scattering (iSCAT)[16] microscopy. iSCAT is a form of inline holography where the reference wave is generated as a back-reflection off an interface, typically the glass/air or glass/solvent interface of the coverglass holding the sample. iSCAT achieves high sensitivity, down to the single protein level[17,18], and allows high-speed observations[19]. Its inline nature makes it readily compatible with fluorescent imaging modalities[20] but also comes with specific drawbacks related to non-trivial signal scaling[21] and twin-image problems[22]. Additionally, backscattering-based approaches struggle when transitioning from the

Rayleigh- to the Mie-scattering regime and parasitic back-reflections generated inside the microscope objective render a conceptually easy-to-implement methodology experimentally very challenging to adopt, especially for larger fields-of-view or when targeting absolute sensitivity limits.

The aforementioned drawbacks can be eliminated by moving away from iSCAT's inline configuration. So-called off-axis holography, or interferometric scattering, also relies on interference between two electric fields but at an angle, that is, in an off-axis configuration[23]. The two fields are typically generated externally which allows choosing appropriate experimental parameters that eliminate twin images, non-trivial signal scaling and parasitic back reflections[24]. While being a popular methodology in the broader optics community highly sensitive, iSCAT-type, nanoscale-measurements are rarely reported. In this tutorial review we will discuss how the concepts of off-axis holography and iSCAT can be seamlessly combined in a highly synergistic fashion to yield quantitative, ambiguity-free, nanoscale observations over extended 3D volumes at sensitivities comparable to iSCAT microscopes.

**Theoretical considerations,**

Interferometric techniques such as iSCAT, inline or off-axis holography rely on the interference between two electric fields. Conceptually, they are all identical. As such, we will refer to the fields involved as signal, $E_s$, and reference, $E_r$, fields, irrespective of the specific technique. The former contains the image information of interest, the latter serves as a reference and is often assumed to carry no additional information. When spatially and temporally overlapping at a detector, these fields interfere thus generating a so-called hologram, $I_{holo}$:

$$I_{holo} = (E_s + E_s^*)(E_r + E_r^*) = E_s^2 + E_r^2 + E_s E_r^* + E_s^* E_r \qquad (1)$$

With $E$ ($E^*$) being the complex (complex conjugate) of the electric field. For sake of simplicity, we have omitted the physical constants ($\frac{1}{2}\epsilon_0 c$) in all equations relating the hologram intensity at the detector and the electric fields. Using $E = Ae^{-i\varphi}$, with $A$ being the electric field amplitude and $\varphi$ its phase, we can rewrite Equation 1 as:

$$I_{holo} = A_s^2 + A_r^2 + A_s A_r e^{-i\Delta\varphi} + A_s A_r e^{+i\Delta\varphi} = A_s^2 + A_r^2 + 2A_s A_r \cos[\Delta\varphi] \qquad (2)$$

Where $\Delta\varphi = (\varphi_{sig} - \varphi_{ref})$ is the phase difference between the signal and reference field. Equation 2 broadly describes all two-field interference experiments. As such, it describes approaches relying on inline-holography such as interference reflection or iSCAT microscopy[25–27]. What makes off-axis holography distinct is its ability to computationally isolate the interference terms, $A_s A_r e^{-i\Delta\varphi}$ or $A_s A_r e^{i\Delta\varphi}$, from the intensity terms, $A_s^2 + A_r^2$, and separate the amplitude and phase information[28,29]. In other words, off-axis holography isolates the signal's complex electric field and unlocks computational post-processing routines that are difficult to combine with inline detection schemes due to the twin image problem[30].

**Off-axis holography: General implementation,**

As the name suggests, off-axis holography relies on interference between non-collinearly traveling signal and reference fields. Figure 1a shows a possible implementation. A beamsplitter generates illumination, $E_{illu}$, and reference, $E_r$, fields, typically from a spatially coherent light source such as a laser. $E_{illu}$ interacts with the sample of interest and a microscope objective collects the signal field, $E_s$, containing illumination light alongside sample-scattering, $E_{sca}$. A lens then propagates $E_s$ onto a camera, placed conjugate with the sample plane, where interference with $E_r$ occurs.

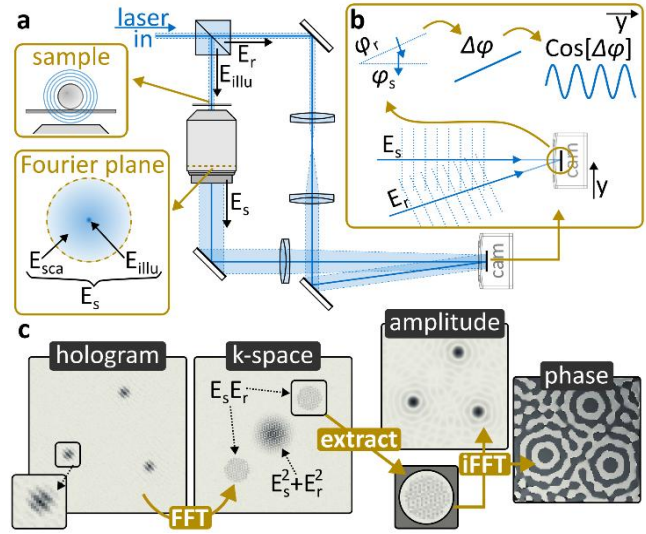

**Figure 1, How-to off-axis holography.** a) A minimum-complexity experimental implementation of off-axis holography. b) Wavefront schematic explaining off-axis induced oscillatory modulation using plane-waves. c) How-to extract phase and amplitude images from a hologram; simulated data. The absolute value of the complex k-space is shown.

The angle between $E_s$ and $E_r$ allows separating the interference and amplitude square terms[29,31,32], as outlined in Figure 1b using plane waves. In brief, the angle between the waves results in a position-dependent linear phase gradient. Assuming an angle in the y-dimension we can rewrite Equation 2 as:

$$I_{holo} = A_s^2 + A_r^2 + 2 A_s A_r \cos[\Delta\varphi_{sample} + ay] \quad (3)$$

With $\Delta\varphi = \Delta\varphi_{sample} + ay$, where $\Delta\varphi_{sample}$ is the sample-induced phase difference between $E_s$ and $E_r$ and $ay$ a linear, y-dependent, phase gradient. Equation 3 shows that the interference term is spatially modulated with $a$. In other words, a Fourier transformation allows isolating it in momentum-, or k-space. Figure 1c summarises the off-axis workflow, from acquired hologram to isolated phase and amplitude images. In a first step, the as-acquired hologram is Fourier transformed into k-space. Here, the amplitude square terms and the interference terms are separated due to the linear phase gradient. The amplitude square terms are located around 0,0 in k-space and two interference terms are visible, a direct result of the complex and complex conjugate (Equation 2) which inverts the phase and with it the off-axis induced phase gradient. Hard-aperture selecting one interference term followed by shifting its centre to 0,0 and inverse Fourier transforming yields the complex interference term in image space which can be separated into its amplitude and phase component.

**Off-axis holography: Magnification and interference angle,**

To successfully implement off-axis holography following the workflow outlined above, it is important to ensure that the interference terms do not overlap with each other or the amplitude square terms. This condition can be satisfied by adequately choosing an image magnification as well as the angle between the k-vectors of $E_s$ and $E_r$. Without going into details, we recommend employing a magnification that ensures that the nominal detector pixel size, $\Delta px$ corresponds to:

$$\Delta px \leq \frac{\lambda}{3.2 NA} \quad (4)$$

, with *NA* being the numerical aperture. Albeit not being the most space-bandwidth efficient implementation, this configuration allows separating all terms along the k-space diagonal thus making it relatively straight-forward to implement. For a detailed discussion, we refer the interested reader to Dardikman *et al.*[33,34] who provide an excellent summary on the topic alongside strategies to improve the space-bandwidth product.

While it is possible to calculate the necessary interference angles[34], we generally determine the correct interference angle experimentally by systematically adjusting it while observing a Fourier transformation of the hologram. Care should be taken to not chose too large angles, a possibility given that aliasing can make a too-large angle indistinguishable from a correct configuration. To avoid this scenario, we initially keep both $E_s$ and $E_r$ in the same horizontal plane and only adjust the vertical plane. Following this first step, we then carefully adjust the horizontal dimension by walking $E_r$ up/down via two adjustable mirror mounts while monitoring the k-space locations of the interference term. This approach allows detecting, and hence avoiding, aliasing. If a diagonally-placed interference term (see k-space in Figure 1c) does not exhibit equal horizontal and vertical displacement then the larger displacement needs to be corrected.

**Off-axis holography: Crucial experimental details,**

The blueprint presented in Figure 1a in principle allows straight-forward implementation of off-axis holography. However, achieving high-quality measurements requires carefully balancing a few crucial experimental parameters, as discussed in detail in Figure 2a. These aspects are related to coherence and wavefront properties that can be non-obvious but need to be accounted for when designing an experiment. From our experience, interferometric stability is generally of no concern in off-axis holography as long as beam heights, integration times and path lengths are kept within reasonable, microscopy-suitable, limits[24]. Gradual, often nanometric, pathlength changes between individual image-acquisitions only result in relative phase shifts which are measured and hence removeable by simply subtracting a constant. As such, we will not discuss these aspects.

**Off-axis holography: Optical path length matching,**

Even for most CW lasers, the optical path length difference between the signal and the reference arm needs to be carefully matched. Figure 2b highlights this aspect by comparing pathlength-difference dependent interferograms recorded for three typical light sources of decreasing temporal coherence length: a diode pumped solid state laser (DPSS, 532 nm, *CW532-100 Roithner Lasertechnik GmbH*), a single-mode laser diode (520 nm, *PD-01298 Lasertack GmbH*) and a frequency doubled femtosecond laser (515 nm, 200 fs, *Pharos Light Conversion*). Around zero delay-difference all sources show satisfactory interference, followed by rapid loss-of-interference. Both the DPSS as well as the diode show surprisingly short coherence lengths, with the diode being almost comparable to the femtosecond source. The DPSS shows a slow beating pattern, whereas, the laser diode shows reoccurring interference maxima at >1 mm delay-intervals. What summarises these observations is that pathlength control is crucial. Mismatches on the <1 mm scale can result in dramatic signal loss, a fact that complicates the experimental setup but comes with an important benefit: coherence gating, which conveniently eliminates parasitic interferences from back reflections from or scattering off optical components. Experimentally, we advise to systematically control the pathlength difference via a manual translation stage and to select the position of maximum interference contrast. This aspect is especially crucial for diode lasers where the distant local interference maxima can show dramatic contrast reduction when compared to the true "time-zero". Generally, the position of maximum interference contrast corresponds to the correctly-matched configuration.

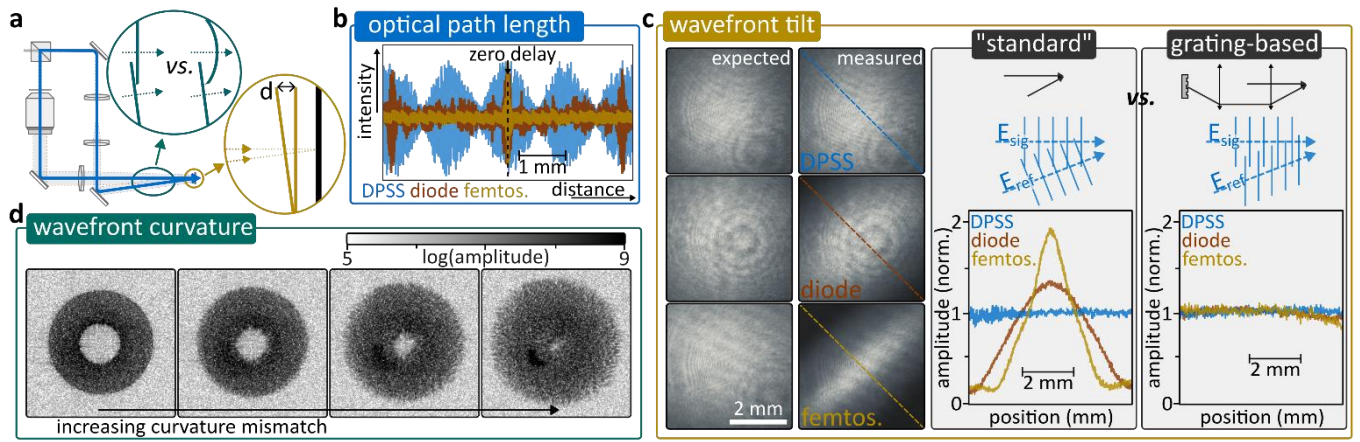

**Figure 2, Experimental details and temporal coherence.** a) Experimental implementation and key points of interest where non-obvious experimental aspects can complicate off-axis holography. b) Interference contrast as a function of optical path length difference between signal and reference arm measured for a DPSS (532 nm, blue line), a single-mode laser diode (520 nm, brown line) and a femtosecond laser (515 nm, yellow line). Zero delay is at the centre of the traces. c) Left: Simulated and measured interference contrast obtained for "standard" off-axis geometry using lasers with varying temporal coherence lengths. The diagonal line marks the direction of maximum angle between signal and reference. Right: comparison of interference-amplitude contrasts for "standard" and grating-based off-axis holography along the diagonal line indicated on the left. d) Impact of wavefront curvature mismatch on the k-space representation of the interferogram.

**Off-axis holography: Adjusting wavefront tilt,**

At short coherence lengths, matching the wavefront tilt at the camera chip can become important, especially for large detectors where signal-to-reference angle induced time-delays easily exceed >100 µm. Figure 2c compares expected and measured interference term amplitudes. The latter are directly obtained via off-axis holography following the workflow outlined in Figure 1. The "expected" amplitudes are computed based on individually measured $A_s^2$ and $A_r^2$ images as the product of the square roots of the two measurements. As can be seen, the short-coherence light sources exhibit reduced amplitudes along the direction of interference, a direct result of the angle-induced pathlength difference. To circumvent this problem, it is possible to generate the reference as the first diffraction order off a grating that is relay imaged onto the camera plane[35,36]. Figure 2c highlights how a grating-based approach allows eliminating loss of interference, a strategy that we successfully used for ultra-broadband fields covering the entire visible spectral range with <2 µm temporal coherence length[37].

**Off-axis holography: Wavefront curvature matching,**

Finally, the wavefront curvature of signal and reference should be matched at the detector plane[38,39]. For infinity corrected objectives plane wave reference fields are often a good starting point. For finite-conjugates focusing the reference at a distance from the detector corresponding to the tube length is a good approximation. Experimentally, we typically transmit the laser through the off-axis setup and then vary the position of a collimation/focusing lens of the reference field until the size of the interference term in k-space is the smallest. We next mount a somewhat concentrated nanoparticle sample onto the microscope to generate a high signal-to-noise ratio projection of the back-focal-plane (BFP) onto the interference term in k-space and then insert a darkfield stop into the objective's BFP. For correctly matched wavefront curvature, the interference term should look like the BFP (Figure 2d). Depending on the level of residual mismatch the BFP is either sharp or defocused. Fine adjustment of the reference curvature, by moving the collimation lens, based on the BFP-appearance allows straight-forward system optimisation. Non-infinity corrected objectives often require a diverging reference wave. Given the low cost of such optics, we advise to simply mirror the microscope in the reference arm to generate a correctly matched reference wave.

**Off-axis holography: Experimental flavours,**

Thus far, we have focused our discussion on image-space holography (Figure 3a) as an intuitive extension of darkfield or iSCAT microscopy. From a work-flow perspective, one optimises the microscope following established routines and then adds holographic capabilities which makes the implementation somewhat straight-forward. Computationally, this modality extracts phase and amplitude information by relying on the position-momentum Fourier relationship. It is therefore also feasible to conduct momentum, or k-space, off-axis holography which ultimately yields image-space images[24]. Figure 3b schematically describes the implementation. Rather than placing the camera into a conjugate image plane, it is placed into a conjugate Fourier plane e.g. at the position of the BFP. Interference is analogous to image-space holography but now a single Fourier transformation is sufficient to isolate the complex image-space interference terms. Experimentally, both approaches have advantages and disadvantages that need to be carefully balanced when selecting for the best implementation.

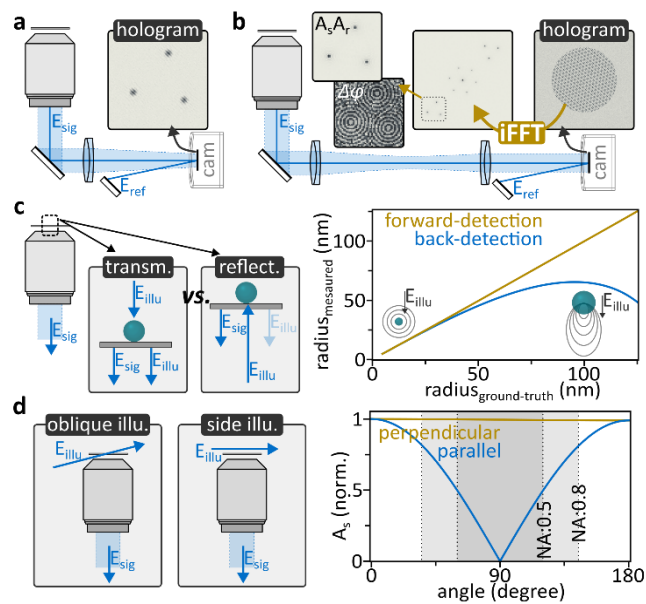

**Figure 3, Experimental flavours.** a) Image-space holography. b) K-space holography. c) Forward vs. back-scattering geometries and particle size-dependent signal scaling. d) Illumination at large k-vectors eliminates parasitic back-reflection and mitigates optic damage but choosing the correct polarisation is crucial. The simulations (c,d) were performed using MiePlot v4.6.21[40,41] assuming a surrounding refractive index of 1.33 and particle refractive index of 1.5 at a wavelength of 532 nm. The polarisation-dependent angular scattering amplitude simulations were performed using the same parameters and a particle radius of 10 nm.

**Off-axis Nanoscopy: Illumination geometries and signal levels,**

Off-axis holography allows freely selecting the illumination geometry which warrants a careful evaluation of angle-dependent scattering amplitudes. Figure 3c,d discuss a few illumination geometries that we commonly use in our labs. Transmission and reflection, e.g. 0 and 180 degree angle of incidence are widely used, corresponding to brightfield and interference reflection microscopy. The former, e.g. forward detection, accurately recovers particle sizes when measured based on scattering amplitudes with $r \propto \sqrt[3]{A}$, whereas the latter rapidly underestimates the size: a direct result of the transition from Rayleigh to Mie scattering (Figure 3c)[24]. This effect leads to size-ambiguities but also has advantages. For instance, it reduces scattering signals of possibly present larger contaminations. Alternative illumination geometries, not applicable to inline-detection, are highlighted in Figure 3d. These implementations do not propagate the illumination light through the microscope objective which eliminates all parasitic reflections and allows dramatically increasing the illumination intensity. Especially the latter feature is highly desirable for large field-of-view observations. Beyond particle size dependent Mie scattering, as discussed previously, the illumination polarisation has to be carefully adjusted when employing oblique- or side-illumination schemes. Figure 3d highlights the dramatic scattering amplitude differences between parallel and perpendicular polarised illumination. Similarly, potential polarisation rotations need to be accounted for to ensure that signal- and reference-waves interfere at the detector. Finally, when employing light sources of short temporal coherence lengths, the position-dependent pathlength-differences for the latter geometries might result in unfavourable signal scaling and need to, hence, be carefully characterised.

**Computational post-processing: z-propagation,**

Following the detailed experimental description we will now focus on the second main pilar of off-axis holography which is a crucial advantage over alternative schemes: computational image post-processing. The capabilities of computational post-processing are unlocked once the complex electric field has been isolated (Figure 1c). For nanosizing applications using particle suspensions, z-propagation, or digital refocussing, allows reconstructing 3D volumes from a single 2D acquisition (Figure 4a)[42–44] as long as the sample is sufficiently sparse. The excellent review by Memmolo *et al.* provides a general introduction to the topic[45]. We implement 3D propagation via the so-called angular spectrum method[46]. In brief, the image-space field is transformed into k-space and then multiplied with the following propagation kernel:

$$K(x, y, z) = e^{\left(-iz\sqrt{k_m^2 - k_x^2 - k_y^2}\right)} \qquad (4)$$

, where $k_m = 2\pi n/\lambda$, with *n* being the refractive index of the propagation medium, *λ* the wavelength and *z* the propagation distance. The discretized spatial frequencies are $(k_x, k_y) = 2\pi(x, y)/(M\Delta x)$ for $(-M/2 \leq x, y \leq M/2)$ with *Δx* being the magnified pixel size of the imaging system after Fourier-extracting the interference term.

This propagation kernel can be understood as a lens function with a wavenumber cut-off determined by the refractive index of the propagation medium which allows reconstructing large 3D volumes. An example is shown in Figure 4b where we computed a 3D image stack from a single plane recording of a 4% agarose gel containing 80 nm diameter Au nanoparticles using a water immersion objective with a numerical aperture of 1.2 (UPLSAPO60XW/1.20, *Olympus*). To evaluate how the z-position of a given particle with respect to the image plane impacts the localisation precision we varied the sample to objective distance by means of a closed-loop piezo, taking three 42 nm steps followed by a larger step. For each step we 3D localised all particles based on volumetric representations as the one shown in Figure 4b. A qualitative comparison between the individual particles' locations reveals that the nanometric steps are detectable for both physically in-focus particles as well as 40 μm out of focus particles (Figure 4c). To further quantify the z-dependent localisation precision we computed the mean position change, using all particle localisations, and compared it to the changes detected on a particle-by-particle level. Figure 4d shows that no significant localisation precision difference is visible for particles located around the physical image plane as compared to far out of focus candidates.

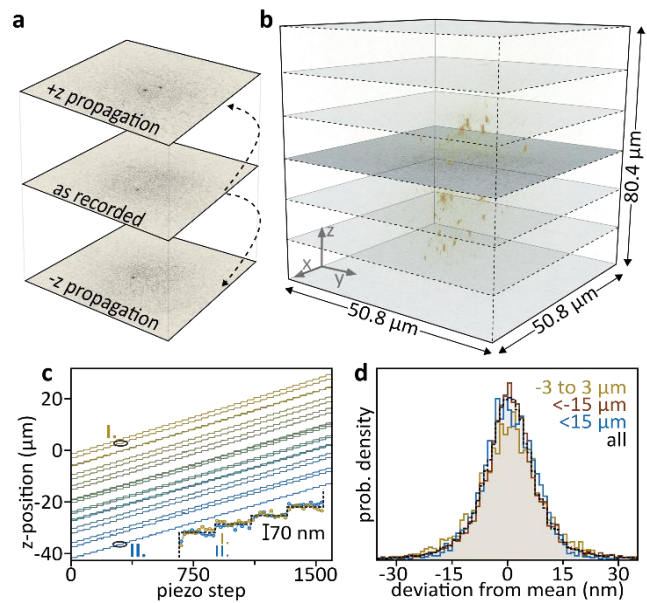

**Figure 4, Numerical propagating.** a) Holographically recorded 2D image planes can be propagated to different z-positions by using a so-called propagation kernel. b) An experimentally acquired hologram (dark plane) is propagated to reveal a 3D volume of the sparse sample composed of 80 nm diameter Au nanoparticles immobilised in a 4% agarose gel (objective: UPLSAPO60XW/1.20, *Olympus*). c) 3D localising the nanoparticles while systematically varying the agarose sample to objective distance by means of a z-piezo shows that precise 3D localisation over extended distances is feasible. Bottom inset: an in focus and 40 µm out of focus nanoparticle show identical piezo-steps. d) Comparison of particle localisations for particles present near the image plane and far out of focus. Essentially identical localisation precisions are obtained thus validating holographic 3D tracking applications.

When propagating over large volumes it is important to keep in mind that signal loss might occur for far out-of-focus particles, which ultimately degrades sensing performance and impacts the recovered scattering amplitudes. The reasons are twofold. First, large defocus means very strong wavefront curvature. As a result, particle-scattering might reach the detector but its interference might be incorrectly detected due to aliasing effects. Second, the scattered light radially spreads which means that some might be lost as it no longer reaches the detector, a common scenario given the limited detector size. Unsurprisingly, this effect is especially severe for objects near the image edge. Combined, both effects might ultimately result in xyz-position dependent amplitude scaling especially for large defocus. Experimentally, a suitable propagation range can be estimated by visually inspecting the resulting xy-images at a given z-position. When reaching the cut-off range effects such as point-spread-function blurring or a sudden drop in observed particle densities, as compared to in focus images, are a clear indication. Importantly, as long as the particles can be localised, the experimentally known parameters such as image size, imaging optics and z-propagation in principle allow re-normalising all scattering amplitudes based on the extracted xyz-positions using a full physical model of the image formation and propagation process.

**Computational post-processing: Aberration correction,**

Defocus is an optical aberration and it should thus come as no surprise that other forms of aberrations such as coma or astigmatism can be computationally corrected for. Compared to the z-propagation discussed above, the challenge is to determine the optical aberrations prior to removing them. Our strategy relies on isolating individual point scatterers to infer pupil-aberrations. An intuitive example using immobilised nanoparticles and a non coverglass-corrected microscope objective is presented in Figure 5. In brief, when the imaging system is used as intended we observed near aberration-free images as shown in Figure 5a. Fourier transforming an image containing only one particle reveals the residual, minimal, pupil aberrations. Upon inserting a slab of glass between the sample and the objective followed by manual refocussing we note a considerably degraded point-spread-function alongside marked spherical pupil aberrations. We remove these aberrations via a two-step approach based on Zernike polynomials. First, we coarsely estimated the aberrations and subtracted them in k-space with the goal being to eliminate the visible phase-wrapping towards high k-vectors which complicates direct fitting approaches. For this initial estimate, we rely on fits using one-dimensional cuts through the centre of the BFP which can be disambiguity-free unwrapped as long as the darkfield-stop region of zero information is ignored. Following subtraction, we obtain wrapping-free pupil aberrations, which are then fitted using the first 21 Zernike polynomials resulting in an essentially aberration-free pupil plane (Figure 5b). Back Fourier transforming into image-space indeed confirms near-perfect aberration correction in line with both experimentally obtained and theoretically expected point-spread-function cross sections (Figure 5c).

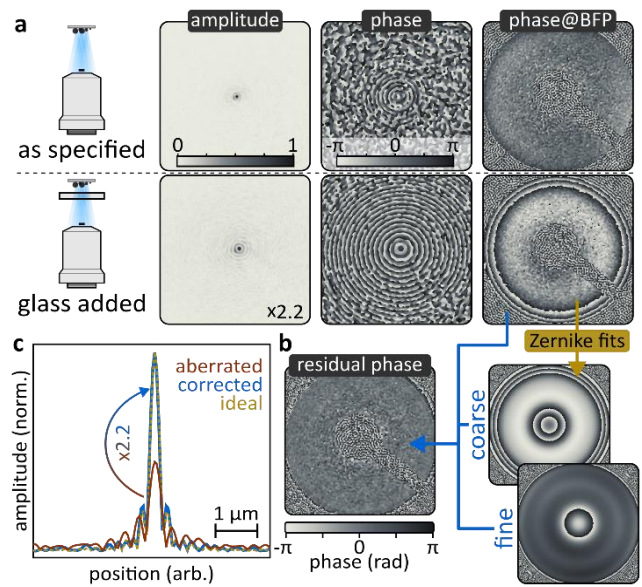

**Figure 5, Aberration correction.** a) Experimentally obtained non-aberrated and intentionally aberrated images of a single Au nanoparticle on glass. The pupil plane of the particle imaged "as specified" indicates a near aberration-free imaging system whereas the same system with glass added between sample and microscope objective shows dramatic pupil aberrations. The missing areas in k-space are due to a darkfield stop. b) Aberration removal based on coarsely estimating the pupil aberrations, to eliminate phase-wrapping, followed by linearly fitting the sum of 21 Zernike polynomials. c) Comparing aberrated, aberration corrected and theoretically expected point-spread-functions demonstrates near-perfect computational aberration correction.

To experimentally implement the aberration correction outlined above on non-ideal, or volumetric, images we typically hard aperture isolated multiple individual nanoparticles in image space. We then set their phase at the centre of the point-spread-function to zero, followed by shifting all particles to the same location, ideally DC. We then average all particles and inverse Fourier transform, followed by the steps outlined in Figure 5. Once the correction pupil phase, $\varphi_{correct}$, is obtained the original hologram is aberration corrected by multiplying its k-space representation by $e^{(-i\varphi_{correct})}$.

**Applications,**

Combined, the steps outlined above allow establishing a working off-axis holographic microscope (Figure 1) and optimising key performance-parameters related to the optical design and light source (Figure 2). Figure 3 explains how to balance illumination parameters and sample choices followed by a summary of key computational post-processing concepts dedicated to z-propagation (Figure 4) and aberration correction (Figure 5). To showcase how these concepts translate to real-world scenarios we conclude with three dedicated applications discussing nanosizing, particle-motion based thermal gradient mapping and the study of photoinduced phenomena.

**Applications: Nanosizing,**

Off-axis holography is ideally suited for size and composition characterisation of synthetic or natural nanomaterials, such as metallic or dielectric nanoparticles, tailored nanometric drug-delivery vectors or extracellular vesicles. At the limits of sensitivity, surface-based inline-holography in a backscattering configuration is arguably the method of choice. It enables extended observation times, which allow achieving sufficient signal-to-noise ratios to detect tiny nano-objects such as single proteins[2,17,18,47]. However, these approaches suffer from important drawbacks. First, the difficulty of separating amplitude square and interference terms can lead to signal ambiguities for particles exhibiting scattering amplitudes comparable to the reference-field amplitude (Equation 2)[24]. Second, when transitioning from the Rayleigh to the Mie scattering regime, scattering signals no longer scale with particle size (Figure 3). Third, as scattering amplitudes are a function of both particle size and composition, it is difficult to distinguish contaminations from analytes of interest. Finally, precise focus control is often necessary, which is costly. Off-axis approaches overcome these drawbacks and do not require costly focus control, making them ideal candidates for commercially viable turn-key instruments for the analysis of unknown or heterogeneous nanoparticle suspensions.

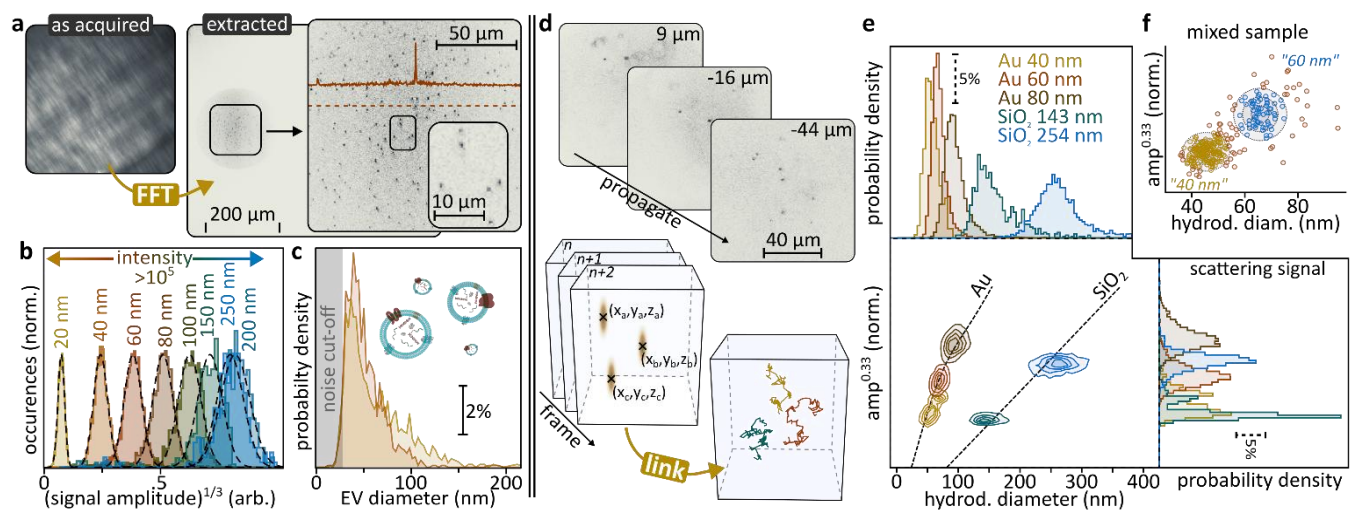

**Figure 6, Nanosizing using off-axis holography.** a) k-space interferogram (left) alongside its Fourier transformation (right) of a sample containing 20 nm diameter Au nanoparticles as observed with a numerical aperture 0.7 microscope objective. b) Scattering signals alongside Gaussian fits (dashed lines) for Au nanoparticles with diameters in the 20-250 nm range. c) Size-distributions of extracellular vesicles samples with a noise cut-off around 25 nm. d) Volumetric 3D particle tracking for advanced nanocharacterisation as enabled

by off-axis holography. e) Holographic nanoparticle tracking analysis (holoNTA) distinguishes nanoparticles based on scattering amplitudes and hydrodynamic diameters. f) holoNTA is well-suited for characterising heterogeneous mixtures. Figures a-c are adapted with permission from *"Precise Nanosizing with High Dynamic Range Holography"* Unai Ortiz-Orruño *et at*. Nano Lett. **21**, 317-322 (2021)[24]. Copyright 2020 American Chemical Society. Figures d-f are adapted from *"Simultaneous Sizing and Refractive Index Analysis of Heterogeneous Nanoparticle Suspensions"* Unai Ortiz-Orruño *et at*. Nano Lett. **17**, 221-229 (2023)[48].

To quantitatively measure heterogeneous clinical nanoformulations we devised k-space holography (Figure 3b and Figure 6a-c),[24] which relies on interference in the back-focal-plane rather than in real space. Compared to real-space imaging, this modality projects the scattering signal of all nanoparticles onto all camera pixels, thus dramatically boosting the achievable dynamic range by approximately six-orders-of-magnitude. A Fourier transformation is sufficient to recover real-space images from the k-space holograms (Figure 6a). Using this approach, we simultaneously measured Au nanoparticles covering a diameter range of 20-250 nm, corresponding to a >$10^5$-fold change in scattering intensity (Figure 6b). The technology enabled directly quantifying SkOV3-derived extracellular vesicle distributions (Figure 6c) based on a low-cost setup, surface-capture and external signal-calibrations using silica nanoparticles.

While powerful, scattering signal-based approaches relying on calibrations are unable to identify the nature of the underlying particles. In other words, a particle of unknown refractive index cannot be correctly sized. This aspect is especially important in the context of extracellular vesicles where it is difficult to distinguish larger protein aggregates from vesicles. To address these limitations, we took advantage of the holographically extended volume of observation which allows robust 3D single particle tracking over long observation times (Figure 6d). Holographic nanoparticle tracking analysis (holoNTA), the combination of scattering-based characterisation with holographically extended 3D nanoparticle-tracking-analysis (NTA) yields two parameters[48,49]. As such, it provides robust particle characterisation: 3D tracking yields hydrodynamic diameters which, combined with scattering signals, allow inferring particle sizes and material composition. Figure 6e highlights the strength of holoNTA when applied to particles exhibiting comparable scattering signals but dramatically differing composition. More specifically, holoNTA was able to distinguish Au and $SiO_2$ nanoparticles of varying sizes based on the two-parameter observations, and also robustly analysed heterogeneous samples (Figure 6f). Compared to near-surface techniques[50,51], holoNTA's extended volumes eliminate the need for high-speed acquisition, which greatly reduces associated equipment cost and, more importantly, eliminates the need for precise nanoparticle localisation[48].

**Applications: Thermal gradient mapping,**

One of the main advantages of off-axis holography versus its inline counterparts is the straightforward single-shot access to quantitative phase information from the sample. Beyond the advantages of propagation and aberration correction already described above, off-axis holography is ideally suited to leverage this property to extract information about the surrounding microenvironment. This is possible because changes to the local microenvironment, in the form of either ionic species[52], temperature[53], or buffer composition[54], give rise to differences in the local refractive index, which in turn can be measured experimentally as relative phase changes. In other words, an off-axis holography microscope operates as a high resolution (diffraction-limited) wavefront sensor, which for instance can be used to reconstruct the 3D thermal gradient landscape.

Figure 7a illustrates the working principle of a holographic temperature gradient sensor, whereby an incident plane wave accumulates an overall phase difference as it travels across a localised refractive index gradient caused by a temperature gradient in the sample volume. Experimentally we generated this thermal gradient by irradiating the sample with a pump beam resonant with the optical absorption of efficient light to heat transducers. In our case, we used plasmonic nanoparticles which allowed us to fabricate an all-optical reconfigurable nano-to-microscale heat source. In this specific example we used a transmission based off-axis system (Figure 3a,c) with a pump-probe scheme to demonstrate how illumination of a single sub-diffraction-limited plasmonic structure (<250 nm) leads to a detectable phase difference. Exploiting established analytical solutions relating temperature fields with measured optical path length differences[53], we retrieved the underlying 3D temperature gradient map.

Notably, because these wavefront-based measurements are intrinsically *in situ*, one can then apply holography-enabled 3D single particle tracking to study how these local perturbations to the microenvironment (e.g. presence of a temperature gradient) affect both single nanoparticle and fluid dynamics. Using tracer beads (1 µm) we were able to capture the dynamics of thermally driven phenomena, such as thermophoresis, convection, and thermoosmosis (Figure 7b). More specifically, the single particle tracking velocimetry approach outlined in Figure 7c allowed decomposing individual particle trajectories into instantaneous 3D displacement vectors (**u**). Ensemble averaging localised vectors over individual voxels allowed extracting high resolution drift velocity ($U_{avg}$) maps alongside the induced temperature gradient (Figure 7d). Using this approach, we identified experimental parameters, which could tune the contribution of each of the thermally driven phenomena to observed dynamics, such as the microchamber height, size and number of heat sources, and orientation of the microchamber with respect to gravity. This nano- to microscale insight related to the respective thermally driven phenomena allows informed engineering of a variety of microfluidic functionalities such as long-range transport[55] or reconfigurable thermal barriers that emulate physical ones[56].

While simple and sensitive, the range of applications for wavefront-based temperature gradient sensors remains limited by the temperature retrieval algorithms. These algorithms are derived from models that assume systems in steady-state with heat sources located in the same plane, and a temperature field smoothly decaying inversely proportional to the distance from the heat sources. These assumptions together with the need of an imaging model relating phase with the temperature fields, can be entirely circumvented by adding a k-vector scanned illumination to the off-axis holographic system, thus converting it into an optical diffraction tomography (ODT) one. The main hallmark of ODT versus off-axis holography is the direct retrieval of the 3D complex refractive index map over the entire imaged volume[57]. Combining pump-probe ODT measurements with a look-up table that relates the refractive index to the temperature of a specific material unlocks time resolving non-steady state (transient) temperature maps without the need of any models[58–60]. Alike off-axis holography, recent advances in ODT have enabled high-speed volumetric tracking of single particles[61].

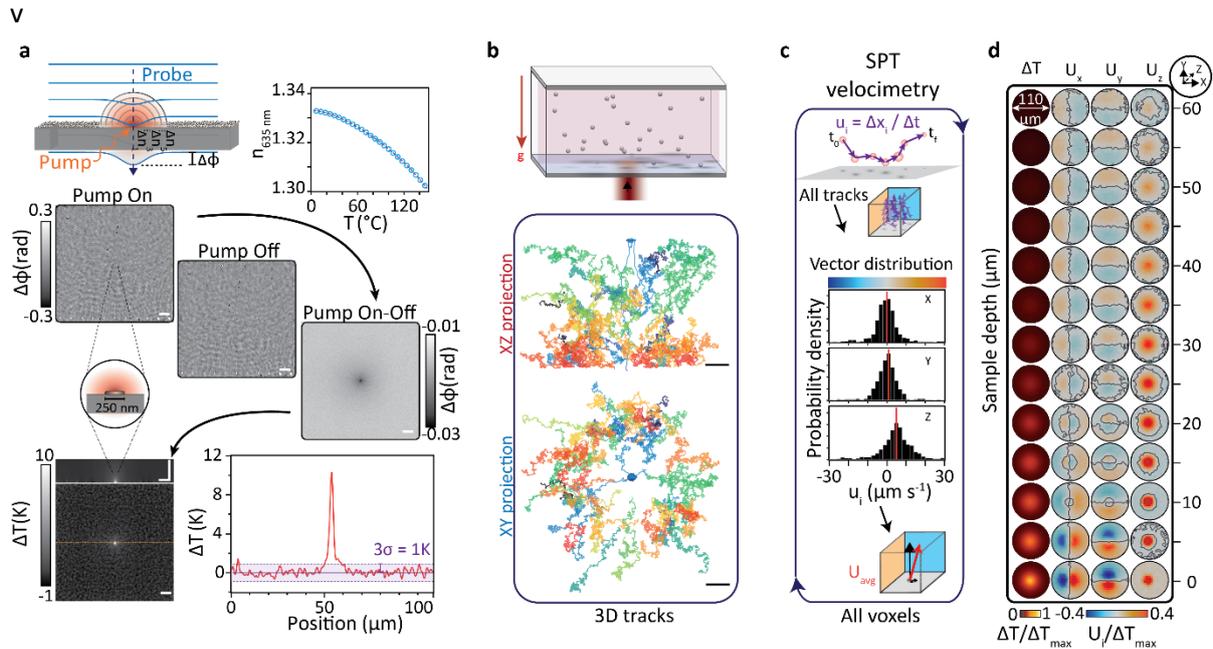

**Figure 7, Thermal gradient mapping using off-axis holography.** a) Working principle of a wavefront-based temperature gradient sensor using a pump-probe approach. b) Representative orthogonal projections of the motion of 1 μm tracer beads in the presence of a thermal gradient contained within a microchamber of nominal height of 50 μm oriented perpendicular to the direction of gravity as shown in the schematic diagram. c) Data analysis workflow to extract drift velocity vectors that capture both particle and fluid dynamics from the 3D tracks of the tracer particles. First, each tracer particle track is segmented into pairwise instantaneous velocity vectors, **u**. Next, the distribution of all instantaneous velocity vectors within a given voxel is computed. Then, from the distribution of **u**, the ensemble average flow velocity vector, **U**$_{avg}$, for each voxel is extracted, thereby suppressing Brownian motion contributions. Finally, this process is repeated for all voxels within the imaged volume. d) Correlative 3D temperature and drift velocity field maps capturing both particle and fluid dynamics upon inducing a thermal gradient inside the microchamber depicted in (b). Figures c-d are adapted with permission from *"Long-range optofluidic control"* Bernard Ciraulo *et al*. *Nat Commun.* **12**, 2001 (2021)[55].

**Applications: Photoinduced changes,**

Off-axis holography allows single-shot multiplexing through the use of multiple illumination and reference waves. These capabilities have been broadly exploited to enable, for example, colour, polarisation or temporal multiplexing[62–65]. Recently, we realised that this concept can be extended to retrieve high-speed signal modulations from long-duration camera exposures thus enabling holographic lock-in widefield imaging[66,67], a modality that had previously been restricted to point-detection. These advances allow visualising rapidly occurring photoinduced processes, at low signal levels, on conventional cameras.

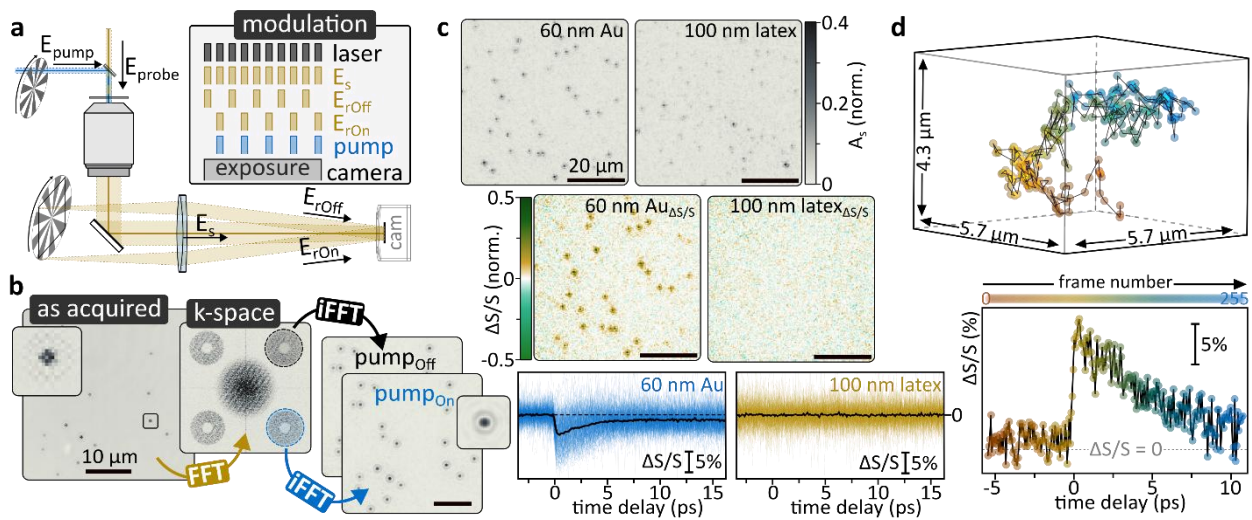

**Figure 8, Monitoring photoinduced dynamics by off-axis holography.** a) Experimental schematic and working principle of all-optical phototransient widefield imaging with a lock-in camera. b) Signal-retrieval scheme for phototransient imaging. c) Phototransient imaging allows distinguishing resonant (Au) and off-resonant (latex) materials based on their photoinduced differential scattering signals. d) Phototransient imaging and pump-probe delay dependent spectroscopy of a freely diffusing 100 nm nanoparticle. Figures a-b are adapted from *"Ultrafast Transient Holographic Microscopy"* Liebel *et at*. Nano Letters **21**, 1666-1671 (2021)[66]. Figures c-d are adapted from *"Widefield phototransient imaging for visualizing 3D motion of resonant particles in scattering environments"* Liebel *et at*. Nanoscale **14**, 3062-3068 (2022)[67].

Figure 8a summarises the working principle of a holographic lock-in camera where a pump-probe experiment is combined with a real-space off-axis holographic microscope employing two reference waves. The probe beam continuously illuminates the sample, while both the pump as well as both reference waves are rapidly modulated. Synchronising pump and reference-wave modulation allows spatially encoding pump$_{ON}$ and pump$_{OFF}$ signals into the same images. Fourier filtering, analogous to single-reference holography, recovers the distinct images from the multiplexed hologram (Figure 8b)[66,67]. When combined with ultrashort pulses, so-called phototransient holography allows studying photoinduced changes on femto- to nanosecond timescales. A distinct advantage over the previously discussed modalities is that phototransient imaging infers additional chemical information through resonant excitations. To demonstrate these capabilities, we conducted initial proof-of-concept experiments on 60 nm Au and 100 nm latex nanoparticles, both of which exhibit comparable scattering amplitudes (Figure 8c). When illuminated with a 400 nm pump, only the Au nanoparticles showed phototransient signal changes, a direct result of hot electron generation via the surface plasmon resonance of Au. By changing the pump-probe time-delay it was, further, possible to follow the nanoparticles' thermalisation dynamics on femto- to picosecond timescales (Figure 8c)[67]. Finally, combining the 3D tracking capabilities with phototransient microscopy allows studying photoinduced dynamics in freely moving objects. Figure 8d shows a 3D trajectory of a freely diffusing 100 nm diameter Au nanoparticle alongside its transient dynamics following photoexcitation.

**Summary and future trends,**

Looking ahead, we envision several exciting avenues for future development. The speed and throughput of GPU-based computation has increased dramatically over the past years, developments which dramatically benefit holographic imaging processing. Their FFT-based nature means that the ever-improving advanced parallelisation schemes immediately expand holographic capabilities both in terms of speed, but also in terms of algorithm-complexities. This development is likely to soon allow real-time 3D hologram analysis and visualisation using desktop-compatible GPUs. These capabilities, in turn, will enable sophisticated data processing and background removal approaches, thus further enhancing off-axis holography's sensing and sizing capabilities. Enabled by these and further technical innovation, we believe that real-time free-flow analysis of even single proteins should be within experimental reach.

By trading ultimate sensitivity for larger volumes of view, via low numerical aperture lenses, it will become possible to observe individual nano-objects for minutes without relying on surface binding. These capabilities will enable real-time studies of nanoscale reactions and photochemistry under relevant experimental conditions, an exciting toolbox that is expected to considerably contribute to the growing insight obtained through so-call operando studies.

Phototransient holography offers exciting opportunities for widefield studying nanoscale photoinduced dynamics in real time. Beyond the currently employed plasmonic systems, tuneable excitation sources will allow applying this promising technology to detecting, analysing and studying dielectric and even biological matter with chemical specificity[68]. We expect contributions in the broader context of widefield photothermal approaches, employing NIR or MIR excitation sources. Here, nanosecond implementations have already uncovered exciting biological phenomena which are difficult to assess with alternative means[60,69]. Expanding such observations to the temporal limits of intramolecular vibrational energy redistribution[70] and nanoscale heat-diffusion[71,72] is likely to contribute valuable insight and will, potentially, allow developing novel diagnostically relevant imaging modalities.

Taken together, we believe that the fusion of holographic imaging modalities with ultrasensitive nanoscopy will facilitate both fundamental studies and enable the development of commercially viable and relevant platforms. This highly promising combination of digital imaging approaches with traditional optics allows replacing costly hardware with exponentially improving *in silico* solutions, and expands the palette of available hardware and software-tools to address the future challenges in characterising heterogeneous nanoparticle samples.